----------
X-Sun-Data-Type: default
X-Sun-Data-Description: default
X-Sun-Data-Name: period.tex
X-Sun-Charset: us-ascii
X-Sun-Content-Lines: 561

\documentstyle[aps]{revtex}
\begin{document}
\preprint{ {\bf /} {\bf /} }
\draft
\title{A method to find unstable periodic orbits for the diamagnetic Kepler problem}
\author{ Zuo-Bing Wu$^{1,2}$ and Jin-Yan Zeng$^1$} 
\address{ Department of Physics, Peking University,
Beijing 100871, China$^1$}
\address{ Laboratory for Nonlinear Mechanics, Institute of Mechanics, Academia Sinica,
Beijing 100080, China$^2$}
\date{\today}
\maketitle

\begin{abstract}
A method to determine the admissibility of symbolic sequences and 
to find the unstable periodic orbits corresponding to allowed symbolic sequences
for the diamagnetic Kepler problem is proposed by using the ordering of stable
and unstable manifolds. By investigating the unstable periodic orbits up to length 6,
a one to one correspondence between the unstable periodic orbits and their
corresponding symbolic sequences is shown
under the system symmetry decomposition.
\end{abstract}

\section{Introduction}

Unstable periodic orbits (UPOs) as the skeleton of chaotic systems
provide a powerful tool to analyze the property of low-dimensional dynamics\cite{Cvitanovic}.
As a simple physical system, the diamagnetic Kepler problem (DKP) 
displays important roles of UPOs in the classical and quantum chaos\cite{THM,WH}.
For the chaotic system, the UPOs are filled in the phase space
and their total number roughly increases with the periodic length. Many UPOs can be found by
searching along three immediate symmetry lines in the configuration space\cite{Eckhardt}.
In view of the scattering of a particle in the four-disk system,
symbolic dynamics has been established\cite{HG} and a method to find UPOs by contracting 
a rectangle in terms of the symbolic dynamics has been given\cite{Hansen2}.
By  considering the stretching and wrapping in the lifted space,
symbolic dynamics has been established\cite{WZ}. In this paper,
we will present a method to find UPOs based on the ordering of stable and unstable manifolds
and describe a one to one correspondence between the UPOs and their
corresponding symbolic sequences for the DKP.
The paper is organized as follows. In Sec. \ref{sec:upo}, we describe 
the method to locate initial values of UPOs in terms of symbolic dynamics.
In Sec. \ref{sec:gen}, by investigating the UPOs up to length 6, 
a one to one correspondence between the UPOs and their
corresponding symbolic sequences is shown under the system symmetry decomposition.
Finally, in Sec. \ref{sec:sum}, some conclusions are given.

\section{Method to locate the initial values of UPOs}
\label{sec:upo}

The classical dynamics of a hydrogen atom with zero angular momentum in a uniform magnetic field $B$ 
along the $z$-axis is described by the ``pseudo'' Hamiltonian (using the semi-parabolic coordinates):
\begin{equation}
h=\frac{p_\mu^2}{2}+\frac{p_\nu^2}{2}-\epsilon(\mu^2+\nu^2)+\frac{1}{8}\mu^2
\nu^2(\mu^2+\nu^2)
\equiv 2,
\label{3}
\end{equation}
where $\epsilon=E \gamma^{-2/3}$ is the scaled energy.
$h$ has $C_{4v}$ and time-reversal symmetry.

In view of the Birkhoff canonical coordinates, a Poincar\'e section
has been chosen along a counterclockwise contour\cite{WZ}.
We choose the $\mu$ and $\nu$ coordinate axes as a Poincar\'e section
and determine a Poincar\'e map in terms of the entering
direction of a orbit. In the first quadrant, the arc coordinate of the map
 is defined as $s=-\mu$ 
along the $\mu$ axis and $s=\nu$ along the $\nu$ axis. 
Using the available transformation $\frac{s}{1+|s|} \to s$,
we get $s \in [-1,1)$. The second coordinate of the map is
$v=\frac{-p_{\mu}}{\sqrt{p_{\mu}^2+p_{\nu}^2}}$ along the 
$\mu$ axis and $v=\frac{p_{\nu}}{\sqrt{p_{\mu}^2+p_{\nu}^2}}$ along the 
$\nu$ axis. According to the $C_4$ symmetry, the 
Poincar\'e maps in the second, third and fourth quadrants can be 
obtained by rotating the map in the first quadrant with angles $\pi/2$, $\pi$
and $3\pi/2$, respectively, around the center anti-clockwisely and 
adding 2,4 and 6 to $s$.
Thus, the dynamics on the Poincar\'e surface is  
represented by the map on the annulus $s\in [-1,7)$ and $v\in [-1,1]$.
The reduced domain  ($s\in [0,2)$, $v\in [-1,1]$) is constructed 
by considering the rotational symmetry. In the same way, the
minimal domain (MD) ($s\in [0,1)$, $v\in [-1,1]$) is taken by
discarding the $\pi$-rotation around the origin. 
 
To illustrate the method to find UPOs, we first discuss the ordering
of stable and unstable manifolds in the Poincar\'e section. In Fig.~1, 
the stable and unstable manifolds in the MD
display the local and global constricting, stretching and 
folding processes of the dynamics. 
When the tangencies associated with folding of one family of the manifolds occur, 
 a partition line through them need to be introduced. 
Using the method given in Ref. \cite{CP}, we roughly determine
the partition line $\bullet C_0$ in Fig.~1.
The demarcation line between two regions with the different rotation numbers
in the lifted space is marked by $\bullet B_0$.
In terms of the natural ordering in the lifted space and the tangencies of manifolds,
we obtain the region partition in the MD with the symbols ($L_0$, $R_0$ and $R_1$)
 and the symbolic ordering of single letter for forward and backward sequences

\begin{equation}
\bullet L_0< \bullet R_0< \bullet R_1,\qquad L_0\bullet <R_0\bullet
<R_1\bullet.
\label{md}
\end{equation}
Except the ordering, the parity of a leading string in the sequences is defined 
by the oddness of the total number of $R_0$. Only odd leading string reverses the ordering.

After the region partition and symbolic ordering are introduced, the families 
of the stable and unstable manifolds make a curvilinear coordinates system in the MD.
All points on a stable (unstable) manifold converge to each other
when iterated forward (backward). In terms of the language of symbolic dynamics,
the points in each stable (unstable) manifold have the same forward (backward) sequence,
i.e., the same ordering. A family of stable (unstable) manifolds is ordered 
according to their transverse intersections with an unstable (stable) manifold.
The ordering is well-defined as long as there are no tangencies
between the two manifolds. The occurrence of tangencies changes
the ordering of folded stable or unstable manifolds, i.e., two families
of sub-manifolds divided by the partition line have the reverse ordering.
In Fig.~1, the ordering of stable (unstable) manifolds increases 
from the left-bottom
(left-top) to the right-top (right-bottom) along each unstable (stable) manifold.
For every point in the MD, there exists a stable or unstable manifold,
which passes through the point and joins with the boundary line $\bullet D_0$.

The ordering of stable and unstable manifolds in the Poincar\'e section
can be displayed by using $\alpha$ and $\beta$ in the symbolic plane.
Every forward sequence $\bullet s_0s_1\cdots s_n\cdots$ may correspond to a 
number $\alpha$, represented in base 3, between 0 and 1.  
The correspondence of symbol $s_i$ to number $\xi_i\in\{0,1,2\}$ is given as $L_0 
\to 0$, $R_0 \to 1$, $R_1 \to 2$ if the leading string $s_0s_1\cdots s_{i-1}$ is even, 
and as $\{L_0,R_0,R_1\} \rightarrow \{2,1,0\}$ for the odd case. Here, the number $\alpha$ is defined as  
 \begin{equation}
 \alpha  = \sum_{i=0}^\infty \xi_i 3^{-(i+1)}.
 \label{eq4} 
 \end{equation} 
Similarly, the number $\beta$ for a backward sequence $\cdots s_{-m}$ $\cdots s_{-2}$ 
$s_{-1} \bullet$ is defined by
 \begin{equation}
 \beta = \sum_{j=1}^\infty \eta_j 3^{-j},
 \label{eq5}
 \end{equation}
where $\eta_j\in\{0,1,2\}$ is determined by $s_{-j}$ and the oddness of 
the leading string $s_{-j+1}\cdots s_{-2}s_{-1}$.
In this way, forward (backward) sequences are ordered according to 
their $\alpha$ ($\beta$)-values.
In particular, at the boundary line $s=0^+$, $\alpha$ ($\beta$)
increases (decreases) monotonically from $v=-1$ to $v=1$.

In Fig.~2, the symbolic plane has a one to one correspondence with the MD.
Many point sets are separated by the boundary line $\bullet D_0$,
the partition line $\bullet C_0$ and their forward or backward maps.
For a  given symbolic sequence, 
its $\alpha_{exact}$ and $\beta_{exact}$ values may be calculated 
by using (3) and (4). 
In the moving process of a starting point, $\alpha_{approx}$ and $\beta_{approx}$
of the current point are respectively determined by using its forward and
backward sequences formed in 50 forward and backward maps of the current point in the MD. 
Our method is described as follows:

(i) We run near $\alpha_{exact}$ from $v$=-1 in the boundary
line $s=0^+$ and 
then iterate $v$ to $\alpha_{exact}$ by using the half-division method. 
In the iteration process, if both $\alpha_{approx}$ and $\beta_{approx}$ for
one of the two points are close to $\alpha_{exact}$ and $\beta_{exact}$
with a given error ($10^{-6}$), we obtain
an initial value of the UPO corresponding to the given symbolic sequence.
The UPO is that passing through the origin.

(ii) If $|\alpha_{approx}-\alpha_{exact}|<10^{-8}$, i.e.,
the difference between the values $\alpha_{approx}$ of two points surrounding $\alpha_{exact}$ is 
less than the error,
we may make a stable manifold from the middle one between the two points. Along the stable manifold,
$\alpha_{approx}$ remains approximately unchanged and $\beta_{approx}$
increases monotonically. When $\beta_{approx}$ is near $\beta_{exact}$, i.e.,
the current point leaps from one side of $\beta_{exact}$ to another,
one should decrease the step size to achieve $|\beta_{approx}-\beta_{exact}|<10^{-8}$. 
However, in this case, the requirement $|\alpha_{approx}-\alpha_{exact}|<10^{-8}$ 
is not met. We may make an unstable
manifold from the middle point and choose a direction according to 
$\alpha_{approx}$. Along the unstable manifold, $\beta_{approx}$
remains approximately unchanged and $\alpha_{approx}$ is monotonically
changed. At this time, if the change of $\beta_{approx}$ or $\alpha_{approx}$
decreases monotonically, the above process will be continued 
until both the requirements $|\alpha_{approx}-\alpha_{exact}|<10^{-15}$ and 
$|\beta_{approx}-\beta_{exact}|<10^{-15}$ are met.
 Thus, in terms of the differences $|\alpha_{approx}-\alpha_{exact}|$ and 
 $|\beta_{approx}-\beta_{exact}|$, 
 we may extract an initial value of the UPO corresponding to the symbolic
sequence, which is allowed. The UPO is that non-passing through the origin.
If the change of $\beta_{approx}$ or $\alpha_{approx}$ does not decrease monotonically
in the 10 circulations, i.e., the algorithm does not converge, the above process will be stopped. 
Thus, we give up the process for $\alpha_{approx}$ and repeat (ii) for $\beta_{approx}$.

(iii) If $|\alpha_{approx}-\alpha_{exact}|>10^{-8}$ 
even when the step of the half-division method is enough small,
i.e., $\alpha_{approx}$ jumps between two nearly invariant values around $\alpha_{exact}$,
we give up the process (ii) for $\alpha_{approx}$ and repeat it for $\beta_{approx}$.

(iv) If $|\beta_{approx}-\beta_{exact}|>10^{-8}$ even when the step of 
the half-division method is enough small,
i.e., $\beta_{approx}$ also jumps between two nearly invariant values around $\beta_{exact}$,
we may say the symbolic sequence is possibly forbidden.

	Using the above method for the given symbolic sequence,
we can obtain an initial point of its corresponding UPO or a possible 
estimation of forbiddenness. For the latter case, we may further repeat 
the above steps to determine the admissibility of symbolic sequences 
produced by shifting the given sequence in its length. We stop the process as soon as  
the symbolic sequence is allowed. If all symbolic sequences generated by shifting
the given sequence in its length are possibly forbidden, we may say the given
symbolic sequence is forbidden.
Finally, we can determine the allowance or
forbiddenness of the given symbolic sequence and extract an initial point of the UPO for
the allowed symbolic sequence. 

To illustrate the main running process of the above method, we take
the symbolic sequences $R_0^2R_1$ and $L_0R_0R_1$ as two examples.
For the sequence $(R_0^2R_1)^\infty \bullet (R_0^2R_1)^\infty$,
using (3) and (4), we get $\alpha_{exact}$=0.5384615384615387 and 
$\beta_{exact}$=0.8461538461538460. The entire process 
to find the UPO is displayed in Fig.~1 and Fig.~2.
First, we begin from the point ($s_0$,$v_0$)=($10^{-10}$,-0.99999) in Fig.~1
and take its symbolic sequence to calculate ($\alpha$,$\beta$) in Fig.~2.
We run along the line $s=s_0$ with a step $\Delta v_0$ and determine the 
last point near $\alpha_{exact}$ by comparing $\alpha_{approx}$ with
$\alpha_{exact}$. Then, we use the half-division method to decrease the difference
between $\alpha_{approx}$ and $\alpha_{exact}$. When the difference is less
than $10^{-8}$, we draw the stable manifold from the point in Fig.~1.
Along the stable manifold, we compare $\beta_{approx}$ with $\beta_{exact}$.
When the difference between them is less than $10^{-8}$, we compare
 $\alpha_{approx}$ with $\alpha_{exact}$ and get their difference
0.0006378404570228. The last point is marked by $a$ in Fig.~3(a).
According to the comparison, we draw the unstable manifold from the point.
When $|\alpha_{approx}-\alpha_{exact}|<10^{-15}$, we compare $\beta_{approx}$ with
$\beta_{exact}$ and get their difference 0.0000029487355183. 
The last point is marked by $b$ in Fig.~3(b). Furthermore,
we continue the above steps with the given error $10^{-15}$. In the second
circulation, when $|\beta_{approx}-\beta_{exact}|<10^{-15}$,
the difference $|\alpha_{approx}-\alpha_{exact}|$ is 0.0000000000441224.
The last point is marked by $c$ in Fig.~3(c);
when $|\alpha_{approx}-\alpha_{exact}|<10^{-15}$, the difference 
$|\beta_{approx}-\beta_{exact}|$ is 0.0000000000000114. 
The last point is marked by $d$ in Fig.~3(d).
In the third circulation, when $|\beta_{approx}-\beta_{exact}|<10^{-15}$, 
the difference $|\alpha_{approx}-\alpha_{exact}|$ is also less than
$10^{-15}$. Thus, we get the point $(s,v)$=(0.6872027097581007,-0.3670108804483804)
marked by $e$ in Fig.~3(d)
corresponding to the sequence $(R_0^2R_1)^\infty \bullet (R_0^2R_1)^\infty$. Also, we may obtain
its forward and backward sequences with 50 letters to conform the result. Finally,
using Newton method, we exactly extract the UPO.
For the sequence $(L_0R_0R_1)^\infty \bullet (L_0R_0R_1)^\infty$,
we get $\alpha_{exact}$=0.1428571428571428 and $\beta_{exact}$=0.8571428571428572.
First, beginning from the point ($s_0$,$v_0$)=($10^{-10}$,-0.99999), 
we run along the line $s=s_0$ with a step $\Delta v_0$ and determine the
last point near $\alpha_{exact}$ by comparing $\alpha_{approx}$ with
$\alpha_{exact}$. Then, using the half-division method, we decrease the difference
between $\alpha_{approx}$ and $\alpha_{exact}$. When $\Delta v_0<10^{-15}$,
the $\alpha_{approx}$ values of the last and current points are 
0.1428571417083783 and 0.1428571738737871, respectively. Since $\Delta \alpha_{approx}>10^{-8}$,
we give up the process for $\alpha_{approx}$ and repeat it for $\beta_{approx}$.
When $\Delta v_0<10^{-15}$, the $\beta_{approx}$ values of the last and current points are
0.8553113553113367 and 0.9046940713607194, respectively. Since $\Delta \alpha_{approx}>10^{-8}$,
the process for $\beta_{approx}$ is stopped. Thus, the sequence $L_0R_0R_1$
is possibly forbidden. Shifting the sequence $L_0R_0R_1$, we get $R_0R_1L_0$ and
extract $\alpha_{exact}$=0.4285714285714285 and $\beta_{exact}$=0.2857142857142856.
We repeat the above process. When $\Delta v_0<10^{-15}$ for $\alpha_{approx}$, 
$\Delta \alpha_{approx}=0.0000000964962270>10^{-8}$. When $\Delta v_0<10^{-15}$ for $\beta_{approx}$,
$\Delta \beta_{approx}=0.0164609053497942>10^{-8}$. The sequence $R_0R_1L_0$
is possibly forbidden. Shifting the sequence $R_0R_1L_0$, we get $R_1L_0R_0$ and
extract $\alpha_{exact}$=0.7142857142857142 and $\beta_{exact}$=0.5714285714285717.
We repeat the above process. When $\Delta v_0<10^{-15}$ for $\alpha_{approx}$, 
$\Delta \alpha_{approx}=0.1481481481481481>10^{-8}$. When $\Delta v_0<10^{-15}$ for $\beta_{approx}$,
$\Delta \beta_{approx}=0.0054869684499316>10^{-8}$. The sequence $R_1L_0R_0$
is possibly forbidden. Since the three sequences generated by shifting $L_0R_0R_1$
are possibly forbidden, the sequence $L_0R_0R_1$ is forbidden.

\section{Correspondence between UPOs and Symbolic Sequences}
\label{sec:gen}

	Before using the method to find UPOs, we discuss qualitatively
the bottom-left region forbidden by the boundary line $\bullet D_0$ in Fig.~2.
Let us focus on the orbits near the origin in the configuration space in Fig.~4. 
Since the orbits are nearly linear, we can 
obtain the relation between two continuous Poincar\'e maps. 
A current point near the origin is described as ($s_0$, $v_0$),
its forward and backward mapping points can be written as ($s_1$,$v_1$)
and ($s_{-1}$,$v_{-1}$), respectively.

(i) If $v_0 \in (-1,0)$, an orbit between the points ($s_0$,$v_0$) and
($s_1$,$v_1$) is nearly linear and has not turning points. We have 
\begin{equation}
\begin{array}{l}
s_1=\frac{s_0 \sqrt{1-v_0^2}}{v_0(s_0-1)+s_0 \sqrt{1-v_0^2}} \approx -s_0 \sqrt{1-v_0^2}/v_0>0, \\
v_1=\sqrt{1-v_0^2},
\end{array}
\end{equation}
where $v_1 \in (0,1)$.
We can obtain the symbolic description $L_0$ for the point ($s_0$,$v_0$) and
$R_0$ or $R_1$ for the point ($s_1$,$v_1$) in Fig.~1. The symbolic string ($L_0^2$)
is forbidden due to the geometry of the orbits. When the coordinate $v_0$ changes
from -1 to 0, the symbolic sequence for the point ($s_0,v_0$) changes from 
$R_1^{\infty}\bullet L_0R_0R_1^{\infty}$
to $R_1^{\infty} R_0\bullet L_0 R_1^{\infty}$ and $\alpha$ ($\beta$) increases
(decreases) monotonically from $\frac{1}{9}$ (1) to $\frac{1}{3}$ ($\frac{1}{3}$). Thus, we get the
left half part of the boundary line $\bullet D_0$ in Fig.~2. 

(ii) If $v_0 \in (0,1)$, an orbit between the points ($s_{-1}$,$v_{-1}$) and
($s_0$,$v_0$) is nearly linear and has not turning points. We have
\begin{equation}
\begin{array}{l}
s_{-1}=\frac{s_0 \sqrt{1-v_0^2}}{v_0(1-s_0)+s_0 \sqrt{1-v_0^2}} \approx s_0 \sqrt{1-v_0^2}/v_0>0, \\
v_{-1}=-\sqrt{1-v_0^2},
\end{array}
\end{equation}
where $v_{-1} \in (-1,0)$.
We can obtain the symbolic description $R_0$ or $R_1$ for the point ($s_0$, $v_0$)
and $L_0$ for the point ($s_{-1}$, $v_{-1}$) in Fig.~1. The symbolic string ($L_0^2$) is 
forbidden due to the geometry of orbits. When the coordinate $v_0$ changes from
0 to 1, the symbolic sequence changes from $R_1^{\infty}L_0\bullet R_0R_1^{\infty}$ to
$R_1^{\infty}R_0L_0\bullet R_1^{\infty}$ and $\alpha$ ($\beta$) increases (decreases) monotonically
from $\frac{1}{3}$ ($\frac{1}{3}$) to 1 ($\frac{1}{9}$). Thus, we get the right half part of the
boundary line $\bullet D_0$ in Fig.~2.

	Therefore, combining (i) and (ii),  
the bottom-left region forbidden by the boundary line $\bullet D_0$ is obtained due to
the linearity of the orbits between two continuous maps
near the origin.

In case of three symbols with the above exclusion rule, the associated connectivity
matrix\cite{E,CE} is

$$\matrix{& & \matrix{ L_0 & R_0 & R_1 } \cr
T=& \matrix{L_0\cr R_0\cr R_1} &
\left(\matrix{
	0	&1	&1\cr
	1	&1	&1\cr
	1	&1	&1\cr}\right)}.$$	
The total number of symbolic sequences of the length $n$ is $N_n=trT^n$,
which can be decomposed into the number $M_d$ of primary symbolic sequences of length 
$d$ dividing $n$: $N_n=\sum_{d|n} dM_d.$ By the M\"obius inversion\cite{HW}, we get 

$$M_n=\frac{1}{n}\sum_{d|n} \mu(\frac{n}{d}) N_d,$$
where $\mu(1)=1$, $\mu(n)=0$ if $n$ contains the square of a prime and $\mu(n)=(-1)^k$
if $n$ contains $k$ prime factors. All the calculated results are shown in Table I.

In Ref. \cite{WZ} we have shown that the symbolic dynamics in the MD removes the rotational 
and reflectional symmetries and only preserves the time-reversal symmetry $T$, which
just reverses the original sequence. We make a sequence set with different lengths
by combining the three symbols. From the set, the symbolic sequences generated by shifting
each primary sequence are removed. If a symbolic sequence generated by the time-reversal
symmetry is only shifts of the primary symbolic sequence, it is also abandoned.
Finally, all symbolic sequences including $L_0^2$ should be removed from the set.
Thus, the number of the sequence set is equal to $M_n$ obtained from
the connectivity matrix, which is shown in Table I.
Under the system symmetry decomposition,
only the time-reversal symmetry is preserved in the sequence set, 
which can be used in cycle expansion\cite{CE}.

Using the method mentioned above, we determine the admissibility of
a sequence set up to length 8 and extract approximate initial points of the UPOs
corresponding to the allowed symbolic sequences for $\epsilon=0$. 
The number of allowed symbolic sequences is presented in Table I.
In the primary sequences, their admissibility as physical UPOs can be further
determined in terms of pruning fronts corresponding to 
the tangencies between stable and unstable manifolds and bound line\cite{WZ}.
For example, in Fig.~5, the points ($\alpha_{\bullet R_0^2R_1}$,
$\beta_{\bullet R_0^2R_1}$) and ($\alpha_{\bullet R_0R_1R_0}$,
$\beta_{\bullet R_0R_1R_0}$) exist in the allowed $\bullet R_0$ region,
and the point ($\alpha_{\bullet R_1R_0^2}$, $\beta_{\bullet R_1R_0^2}$)
exists in the allowed $\bullet R_1$ region. So, the sequence
$(R_0^2R_1)^\infty \bullet (R_0^2R_1)^\infty$ is allowed.
The point ($\alpha_{\bullet L_0R_0R_1}$, $\beta_{\bullet L_0R_0R_1}$)
does not exist in the allowed $\bullet L_0$ or $\bullet R_0$ or
$\bullet R_1$ region, but in the forbidden region of boundary line $\bullet D_0$.
So, the sequence $(L_0R_0R_1)^\infty \bullet (L_0R_0R_1)^\infty$ is forbidden.
We use the method to determine the admissibility of primary sequences up to length 8.
The result is consistent with the above one. 

For a real number,
our Sun-workstation has significant digits with the 16's space.
Due to the restriction of precision, the exact UPO up to length 6 can be further
exacted by using the Newton method.
In order to check a one to one correspondence between
the UPOs and their corresponding symbolic sequences,
we display the allowed symbolic sequences up to period 6 in Table II and
their some corresponding UPOs in Fig.~6.
 In terms of the rotational and reflectional symmetries, 
the Orbit Period is the same or twice or quadruple the Sequence Period,
which can be determined in the configuration space. 
In the 38 UPOs of Table II, except the pairs of UPOs (5) and (6), (11) and (15),
(14) and (16), (25) and (29), (26) and (32), (27) and (33) shown in Fig.~6, other UPOs have the different configurations
and cannot be produced from one to another in terms of the time reversal.
For each one of the 26 UPOs, the time-reversal orbit is the same as that produced by the rotational
symmetry, or reflectional symmetry, or their combination. So, the time-reversal symmetry $T$ 
is degenerate, i.e.,
the UPO and its time reversal orbit correspond to the same symbolic sequences in Table II. 
For the pairs of UPOs (11) and (15), (14) and (16), (26) and (32), (27) and (33), 
they have also the different configurations, but the two UPOs of each pair have 
the same one.
For each one of the two orbits, its time-reversal orbit is another one, which cannot
 be produced from the given UPO by the rotational symmetry, or reflectional symmetry, 
 or their combination. So, the time-reversal symmetry $T$ is non-degenerate.
The two orbits with the non-degenerate $T$ symmetry correspond to the different 
symbolic sequences in Table II.
For the pairs of UPOs (5) and (6), (25) and (29) passing through the origin, 
they have also the different configurations, but the two UPOs of each pair have
the same one even when the time reversal symmetry is considered.
When the right limit $s=0^+$ of the UPOs is taken into account in the configuration space, 
the intersection points of the UPOs (5) and (29), (6) and (25) should be removed from the origin into 
the third and first quadrants, respectively.
At this time, for the pair of UPOs (5) and (6) with the right limit,
the two lines of the UPO in the center should be removed to pass through the axes $\nu>0$ and $\nu<0$, respectively; 
for the pairing UPOs (25) and (29)
with the right limit, the two lines of the UPO in the center should be removed to pass through the axis $\nu>0$. 
The pairs of UPOs (5) and (6), (25) and (29)
with the right limit have the similar configurations to the above UPOs
no-passing through the origin and the same results can be concluded. 
Thus, a one to one correspondence between
the UPOs and their corresponding symbolic sequences is shown under the system
symmetry decomposition. 

\section{Conclusion}
\label{sec:sum}
According to the ordering of stable and unstable manifolds, we propose
a method to determine the admissibility of symbolic sequences and 
to find UPOs corresponding to allowed symbolic sequences
for the diamagnetic Kepler problem. By searching the UPOs
up to length 6,
a one to one correspondence between the UPOs and their
corresponding symbolic sequences has been shown under the system symmetry decomposition.

\acknowledgments
{ This work was supported in part by Post-Doctoral Foundation 
and Nonlinear Science Project of China (ZBW) and
National Natural Scientific Foundation of China (JYZ).}



\begin{table}
\label{tab1}
\begin{footnotesize}
Table I. Numbers of the total sequences (TS), the primary sequences (PS) and the allowed sequences (AS).
\begin{tabular}{llllllllll}
&$n$	&1 	&2	&3	&4	&5	&6	&7	&8\\
TS&$N_n$	&2	&8      &20     &56     &152    &416	&1136	&3104\\
PS&$M_n$&2	&3	&6	&12	&30	&65	&162	&381\\
AS&$\epsilon=0$   &1	&2	&1	&5	&12	&17	&39	&78\\
\end{tabular}
\end{footnotesize}
\end{table}

\begin{table}[htbp]
\label{tab2}
\begin{footnotesize}
Table II. The allowed sequences encoding UPOs for $\epsilon=0$. By $O_i$ we mean
the $i$-th orbit according to the numbering in the first column. 
The Sequence Period and Orbit Period are defined by the length of 
symbolic sequences and periodic points of UPOs in the Poincar\'e section, respectively.
\vspace{0.5cm}
\begin{tabular}{rclcc}
No.	& Sequence Period	& Sequence		& Orbit Period& Non-degeneracy Symmetry\\
\tableline
1	& 1  	& $R_0$ 		&4 	&\\
2	& 2  	& $L_0R_1$ 		&4 	&\\
3	& 2  	& $R_0R_1$ 		&4 	&\\
4	& 3  	& $R_0^2R_1$ 		&6 	&\\
5	& 4  	& $L_0R_0^2R_1$ 	&8 	&\\
6	& 4  	& $L_0R_1R_0^2$ 	&8 	&	$T\circ O_5$\\
7	& 4  	& $L_0R_1R_0R_1$ 	&4 	&\\
8	& 4  	& $R_0^3R_1$ 		&8 	&\\
9	& 4  	& $R_0^2R_1^2$ 		&8 	&\\
10	& 5	& $L_0R_0L_0R_1^2$	&20	&\\
11	& 5	& $L_0R_0^2R_1^2$	&20	&\\
12	& 5	& $L_0R_1L_0R_1^2$	&10	&\\
13	& 5	& $L_0R_1R_0^2R_1$	&20	&\\
14	& 5	& $L_0R_1R_0R_1^2$	&10	&\\
15	& 5	& $L_0R_1^2R_0^2$	&20	&	$T\circ O_{11}$\\
16	& 5	& $L_0R_1^2R_0R_1$	&10	&	$T\circ O_{14}$\\
17	& 5	& $R_0^4R_1$		&10	&\\
18	& 5	& $R_0^3R_1^2$		&20	&\\
19	& 5	& $R_0^2R_1R_0R_1$	&20	&\\
20	& 5	& $R_0^2R_1^3$		&10	&\\
21	& 5	& $R_0R_1R_0R_1^2$	&10	&\\
22	& 6	& $L_0R_0L_0R_1R_0R_1$	&12	&\\
23	& 6	& $L_0R_0L_0R_1^3$	&12	&\\
24	& 6	& $L_0R_0^2L_0R_1^2$	&6	&\\
25	& 6	& $L_0R_0^4R_1$		&12	&\\
26	& 6	& $L_0R_0^3R_1^2$	&6	&\\
27	& 6	& $L_0R_0^2R_1^3$	&12	&\\
28	& 6	& $L_0R_1L_0R_1R_0R_1$	&12	&\\
29	& 6	& $L_0R_1R_0^4$		&12	&	$T\circ O_{25}$\\
30	& 6	& $L_0R_1R_0^3R_1$	&12	&\\
31	& 6	& $L_0R_1R_0R_1R_0R_1$	&12	&\\
32	& 6	& $L_0R_1^2R_0^3$	&6	&	$T\circ O_{26}$\\
33	& 6	& $L_0R_1^3R_0^2$	&12	&	$T\circ O_{27}$\\
34	& 6	& $R_0^5R_1$		&12	&\\
35	& 6	& $R_0^4R_1^2$		&12	&\\
36	& 6	& $R_0^3R_1R_0R_1$	&12	&\\
37	& 6	& $R_0^3R_1^3$		&12	&\\
38	& 6	& $R_0^2R_1^4$		&12	&\\
\end{tabular}
\end{footnotesize}
\end{table}

FIGURE CAPTION
\figure{Fig.~1 Stable and unstable manifolds with two partition lines
in the MD. Circles show the entire process to find the point corresponding
to the sequence $(R_0^2R_1)^\infty \bullet (R_0^2R_1)^\infty$ in the MD.
\label{fig1}}
\figure{Fig.~2 A symbolic plane corresponding to the MD. Circles show
the entire process to find the point corresponding to the sequence
$(R_0^2R_1)^\infty \bullet (R_0^2R_1)^\infty$ in the symbolic plane.
\label{fig2}}
\figure{Fig.~3 (a) A local sketch in Fig. 1; (b) A local sketch in (a);
(c) A local sketch in (b); (d) A local sketch in (c). 
\label{fig3}}
\figure{Fig.~4 A chaotic orbit displaying its linearity near the origin. 
\label{fig4}}
\figure{Fig.~5 Determining the allowance of sequence 
$(R_0^2R_1)^\infty \bullet (R_0^2R_1)^\infty$ and the forbiddenness of
sequence $(L_0R_0R_1)^\infty \bullet (L_0R_0R_1)^\infty$ in the symbolic
plane by pruning fronts.
\label{fig5}}
\figure{Fig.~6 Some UPOs corresponding to the symbolic sequences in Table II. 
\label{fig6}}
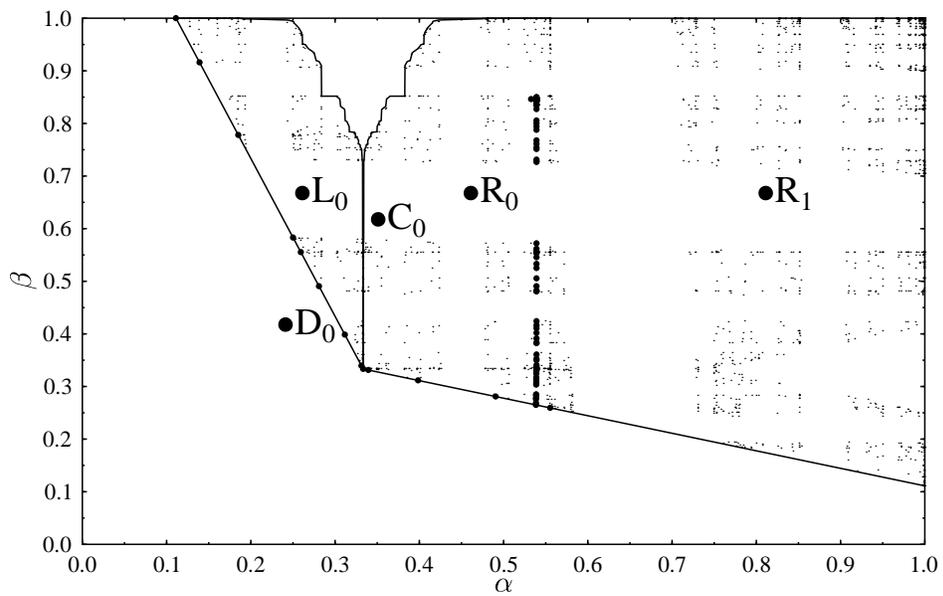
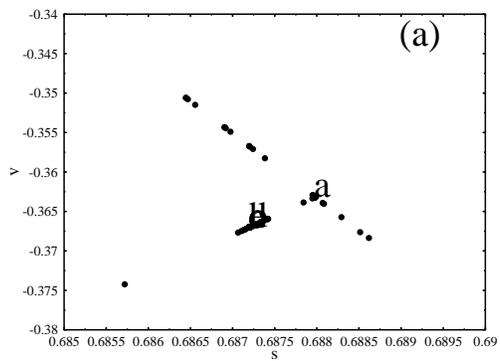
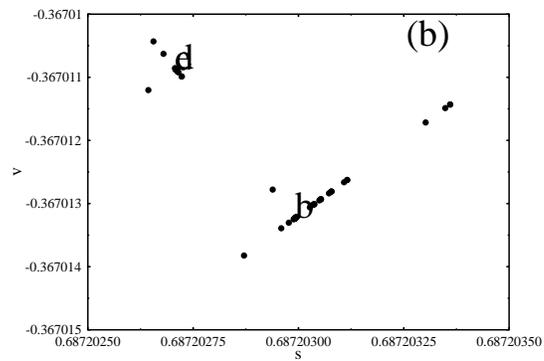
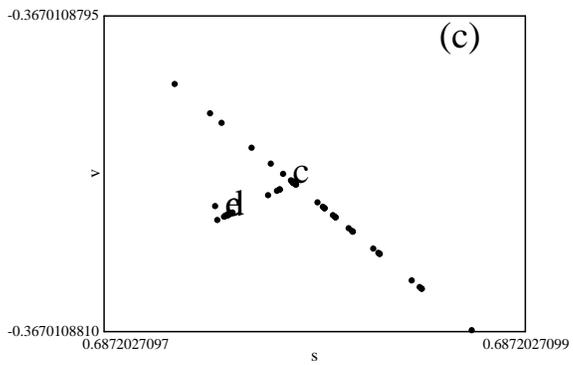
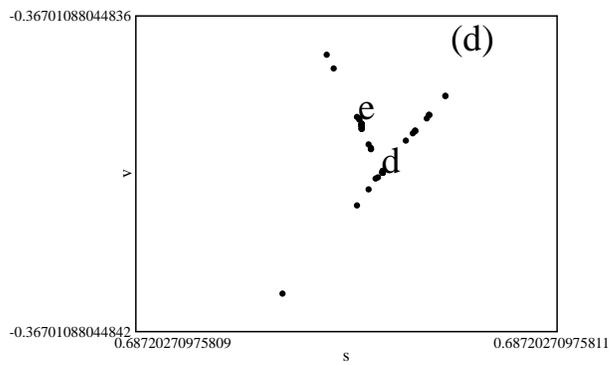
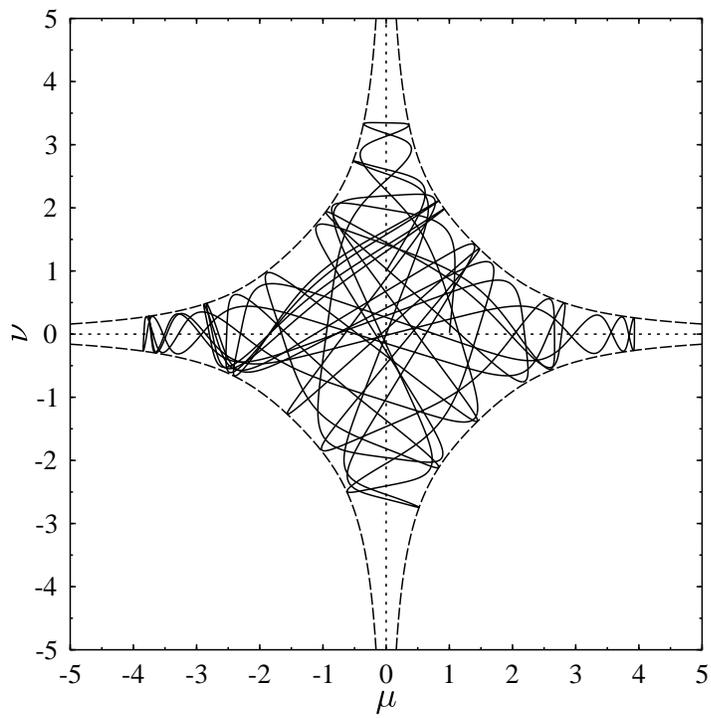
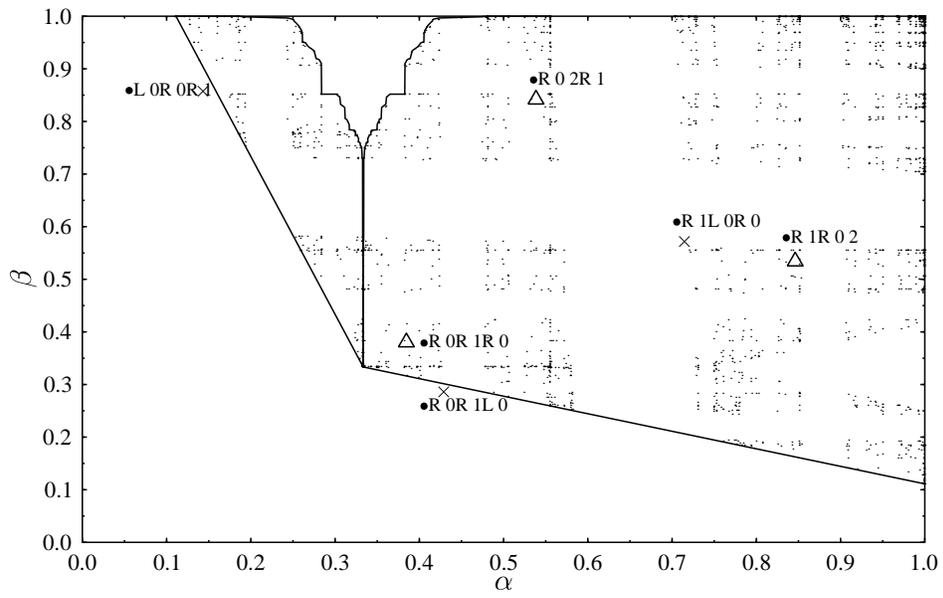
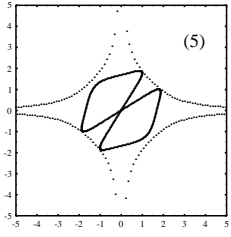
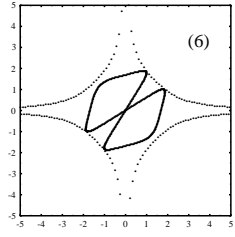
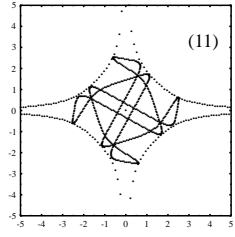
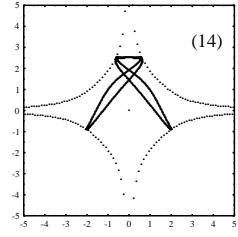
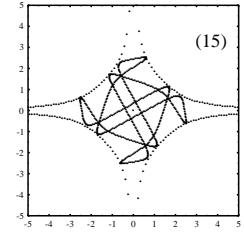
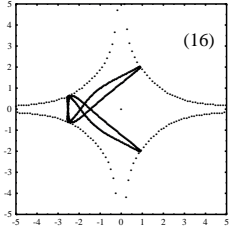
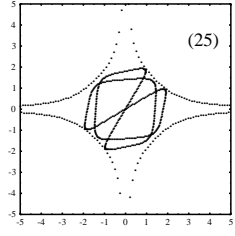
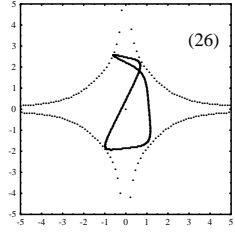
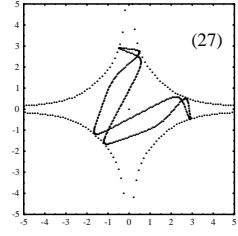
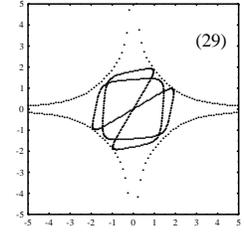
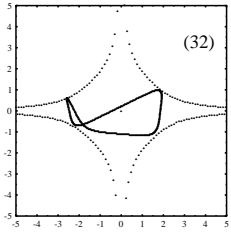
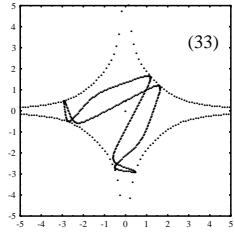
\end{document}